\documentstyle{mn}
\input epsf
\def\plotone#1{\centering \leavevmode
\epsfxsize=\columnwidth \epsfbox{#1}}

\def\apj{ApJ}                 
\def\apjl{ApJ}                
\def\apjs{ApJS}

\def\aap{A\&A}

\def\mnras{MNRAS}

\def\pasj{PASJ}

\def\nat{Nature}

\title[X--ray time lags in the black hole candidates]{On the X--ray
time lags in the black hole candidates}

\author[Kotov, Churazov \& Gilfanov]{O.~Kotov,$^{2,1}$
E.~Churazov,$^{1,2}$ M.~Gilfanov,$^{1,2}$\\
$^1$ MPI fur Astrophysik, Karl-Schwarzschild-Strasse 1, 85741
Garching, Germany \\
$^2$ Space Research Institute (IKI), Profsouznaya 84/32, Moscow 117810, 
Russia}

\pagerange{\pageref{firstpage}--\pageref{lastpage}}
\pubyear{2001}

\begin{document}
\maketitle

\label{firstpage}
\begin{abstract}
	It is shown that the energy dependence of the time lags
in Cygnus X-1 excludes any significant
contribution of the standard reflected component to the observed
lags. The conclusion 
is valid in the 0.1--10 Hz frequency range where time lags have been
detected with high enough significance. In fact the data hint 
that reflected component is working in opposite direction, reducing
the lags at energies where contribution of the reflected component is
significant. 

We argue that the observed logarithmic dependence of time lags on
energy can be understood as due to the small variations of the
spectrum power law index in a very simple phenomenological model. We
assume that an optically thin flow/corona,  emitting a power law like
spectrum, is present at a range of distances from the compact
object. The slope of the locally emitted spectrum is a function of
distance with the hardest spectrum emitted in the innermost
region. If perturbations with different time scales are
introduced to the accretion flow at different radii the observed
X--ray lags naturally appear due to the inward propagation of 
perturbations on the diffusion time scales.   
\end{abstract}


%

\sloppypar

\section{Introduction}
Strong variability of the X--ray flux is a generic property of the black
hole binaries in the so called hard state (see e.g. van der Klis
1994). While it is usually assumed that X--ray emission is produced
via Comptonization of the low frequency photons by an optically thin cloud
of hot electrons (e.g. Sunyaev and Truemper 1979) the nature of the strong
variations is less well understood. One of the possible clues to the
origin of the variability may be connected with the presence of the
time lags between the light curves in different energy
bands. The time lags were first found in the early observations of Cygnus
X-1 (Priedhorsky et al. 1979, Nolan et al. 1981) and confirmed in the
numerous subsequent observations of Cygnus X-1 and other black hole
candidates. In the vast majority of cases hard photons lag with respect to
the soft ones. From observations it was found that (i) the value of the
time lag depends on frequency, (ii) time lag approximately
logarithmically depends on the energy separation between two bands (e.g.
Miyamoto \& Kitamoto 1989), (iii) cross-correlation function between two
bands peaks at zero lags within $\sim$1 ms (Maccarone et al. 2000 and
references therein). Comprehensive description of the observed
Cygnus X-1 variability and time lags are given by Nowak et al., 1999a,b
and Pottschmidt et al. 2000. A recent reviews of the observed lags
and theoretical models are given by Poutanen 2001a,b.

We discuss below the time lags using RXTE data on Cygnus X-1 as the example.
In Section 2 we show that observed energy dependence of the time lags
excludes significant contribution of the standard reflected component to the
time lags in Cygnus X-1. In Section 3 we
introduce simple phenomenological model of the time lags, which naturally
explains the logarithmic energy dependence of the time lags. In
Section 4 we discuss the possibility of explaining the lags as due to
diffusive propagation of the perturbations in the accretion flow. In
Section 5 we discuss implications of the results. Section 6 summarizes
our finding. 

\section{Time lags due to light crossing time of the reflector}
For hard X--ray state of Cygnus X-1 one of the popular models
assumes that an optically thick and geometrically thin disk is
truncated at some radius from the compact object and most of the
observed X--ray emission is coming from the more compact optically
thin region located closer to the black hole (e.g. Esin,
McClintock and Narayan 1997, Zdziarski et al. 1999). A
fraction of the hard X--ray photons emitted by the central region can
be intercepted by the optically thick disk and reemitted again. This
re-emitted/reprocessed component, also called ``reflected'' component,
has a very distinct spectral shape (e.g. Basko, Sunyaev \& Titarchuk 1974,
George \& Fabian 1991). If matter in the
optically thick disk is neutral or weakly ionized than the spectrum of 
the reflected component is harder (in the standard X--ray band) than
the illuminating spectrum and it also contains prominent iron 
fluorescent $K_\alpha$ line. Because of the delay of photons which
traveled from the emission region to the reflector and then to the
observer hard lags may naturally appear (e.g. Poutanen 2001b).

In the presence of the direct and reflected components the observed spectrum
$S(E,t)$ at time $t$ can be written as:
\begin{eqnarray}
S(E,t)=D(E,t)+R(E,t)
\end{eqnarray}
where $D(E,t)$ and $R(E,t)$ are the time dependent spectra of the direct and
reflected components respectively. 
Assuming that the direct component vary in intensity only and not in
shape: 
\begin{eqnarray} 
D(E,t)=A(t)D(E)
\end{eqnarray}
and neglecting the dependence of
the reflected component spectral shape on the incident and emission
angles the reflected component can be written as:
\begin{eqnarray} 
R(E,t)=\frac{\Omega}{2\pi} R(E) \int T(\tau) A(t-\tau) d \tau 
\label{conv}
\end{eqnarray}
where $R(E)$ are the energy spectra of the reflected component calculated
for the reflector subtending solid angle of $2\pi$, $A(t)$ is the
normalization of the direct component and $T(t)$ is the reflected component
transfer function, which characterizes the time dependence of the intensity of
the reflected component in response for a infinitely short flare of the 
direct component. In the above expression $T(t)$ is normalized such
that $\int_0^\infty T(t) dt=1$ and the contribution of the reflected
component to the time averaged spectrum is expressed through the
standard factor $\frac{\Omega}{2\pi}$, where $\Omega$ is the total solid
angle subtended by the reflector.\\
The expression (1) can be rewritten as:
\begin{eqnarray}
S(E,t)=A(t)D(E)+\frac{\Omega}{2\pi} R(E) \int T(\tau) A(t-\tau) d \tau   
\label{set}
\end{eqnarray}

Further assuming that contribution of the reflected component to the total
spectrum is small one can write the expression for a frequency resolved
spectrum, introduced in Revnivtsev et al. 1999:
\begin{eqnarray} 
S(E,f)=|\hat S(E,f)|\approx \nonumber\\
|\hat A(f)| D(E)\times
\sqrt{1+2\frac{\Omega}{2\pi}\frac{R(E)}{D(E)}Re[\hat T(f)]} \approx
\nonumber\\ 
|\hat A(f)|\times (D(E)+\frac{\Omega}{2\pi}R(E)Re[\hat T(f)])
\end{eqnarray}
The equivalent width of the iron line in the frequency resolved spectrum is
then obviously:
\begin{eqnarray} 
EW(f) \approx EW_t \times Re[\hat T(f)],
\label{ewcor}
\end{eqnarray}
where $EW_t$ is the equivalent width in the total source spectrum.
From the above expression it is clear that the equivalent width in the
frequency resolved spectrum should decrease if the transfer function is
smooth at a given frequency. E.g. when the light crossing time of the
reflector is large the equivalent width should be low at the corresponding
time scales. Using these arguments Revnivtsev et al. 1999, Gilfanov et al.
2000 set an upper limit on the effective size of the reflector in Cygnus
X-1 of $\sim$50 $R_g$. We note here that the above expression assumes linear
relation between variabilities in the reflected and direct components in the
form of the convolution (\ref{conv}). If on the contrary the variability of the
direct and reflected components are completely incoherent/uncorrelated then
the contribution of the reflected component to the frequency resolved
spectrum is proportional to square of the solid angle subtended by the
reflector $(\frac{\Omega}{2\pi})^2$.

\begin{figure} 
\plotone{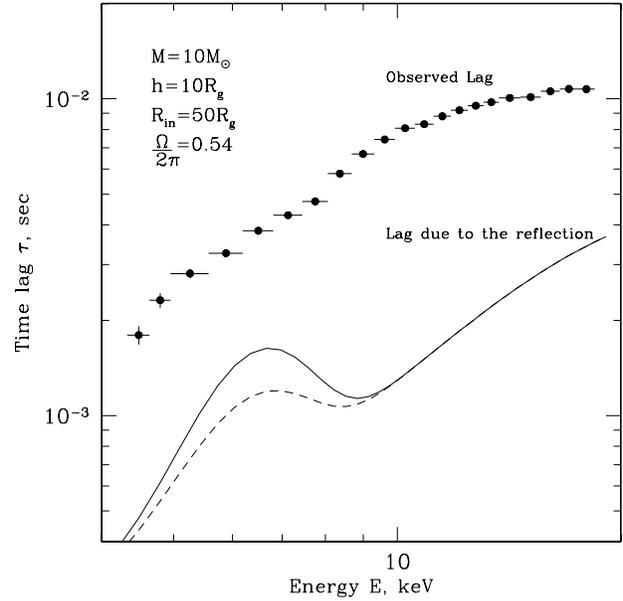}
\caption{The energy dependence of the time lag in Cygnus X-1 in the
hard state at the frequency of 2.5 Hz. For comparison the energy dependence
of the time lags expected in the model with extended reflector is shown. For
upper curve we used relative normalization of the Gaussian and reflected
continuum obtained from the fit to the total Cygnus X-1 spectrum during
observation 10238-01-08-00 (Gilfanov et al., 1999). For the lower curve 
the normalization of the Gaussian was chosen such that equivalent width of
the line with respect to the reflected continuum is $\sim$1 keV. The latter
value is characteristic for an angle averaged reflection from a neutral
matter with a normal abundance of heavy elements. 
\label{lag}
}
\end{figure}

\begin{figure}
\plotone{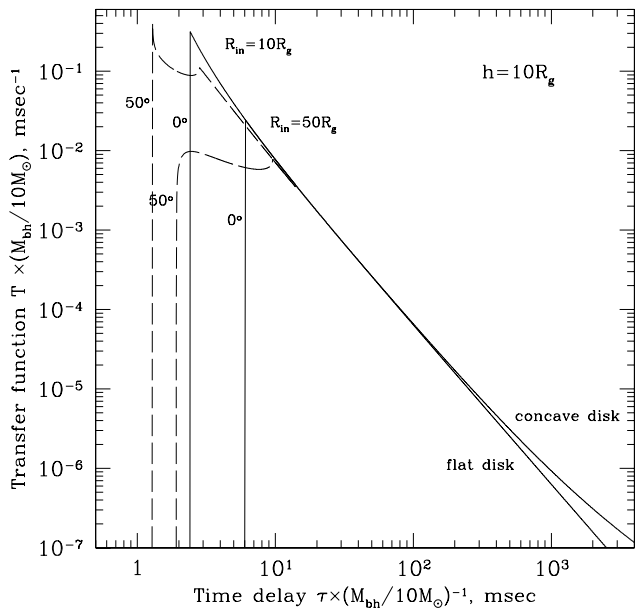}
\caption{Transfer functions $\frac{\Omega}{2\pi} T(t)$ for an isotropic
point source of primary radiation located at $h=10R_g$ above a flat or
concave disks  with inner  an radius of $R_{in}=10$ and $50 R_g$  and
inclination of $0^\circ,50^\circ$. 
\label{transfer}
} 
\end{figure}

One can further test the hypothesis of the large effective size of the
reflector using the time lags. Assuming again that the light curve in any
energy band can be expressed through (\ref{set}) the expected phase
lag $\phi(E,f)$ between given energy $E$ and much lower energy (where
contribution of the 
reflected component is negligible and the spectrum is solely due to direct
component) can be written as: 
\begin{eqnarray}
tg(\phi(E,f))=\frac{Im[\hat S^*(E,f) \hat D(E,f)]}{Re[\hat S^*(E,f) \hat
D(E,f)]}\approx\nonumber\\  
\frac{\frac{\Omega}{2\pi}R(E)
Im[\hat{T}]}{D(E)+\frac{\Omega}{2\pi}R(E) 
Re[\hat{T}]}\approx \frac{\Omega}{2\pi}\frac{R(E)}{D(E)} Im[\hat T (f)]
\label{eq:refl}
\end{eqnarray}

 Therefore in this approximation the frequency dependence of the phase/time
lags is due to imaginary part of the Fourier transform of the transfer
function. Energy dependence of the lags is simply the ratio of the reflected
and direct spectral components. Since $D(E)$ is assumed to be smooth
function of energy the time/phase lag as the function of energy should
contain prominent iron line with the equivalent width comparable with the
equivalent width of the line in the reflected component, i.e. much larger
equivalent width than in the total spectrum. 

\begin{figure}
\plotone{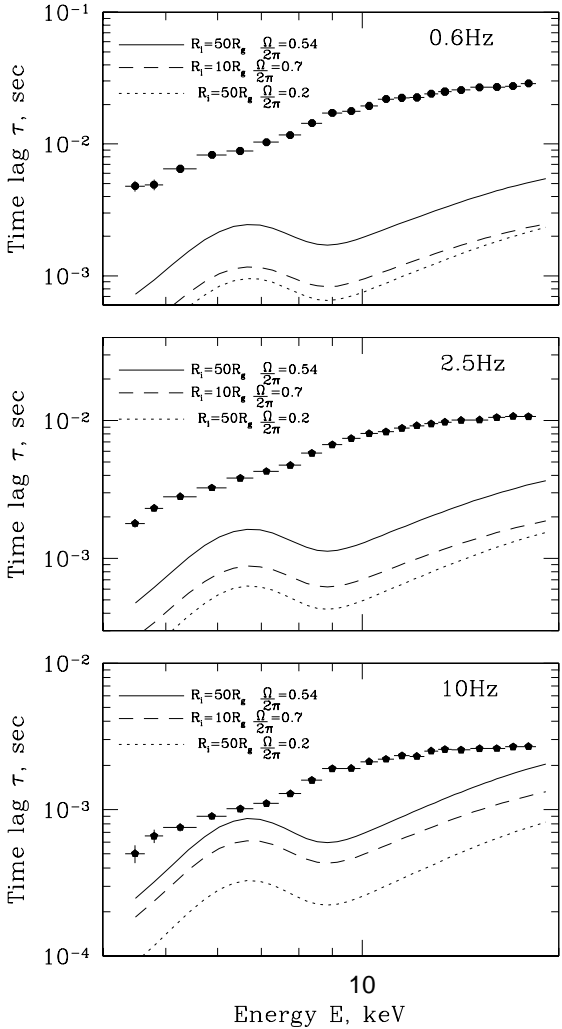}
\caption{The energy dependence of the time lag in Cygnus X-1 in the
hard state at different frequencies. For comparison we show the expected
curves for the models where reflected component originates from a flat disk
with a central hole illuminated by the compact source located at the height
$h=10 R_g$ above the disk surface. The model curves were
calculated for the equivalent width of the Fe $K_{\alpha}$ line obtained
from the fit of the time averaged source spectrum. 
\label{compar}
}
\end{figure}

In order to assess possible contribution of the reflected component in
the observed time lags in Cygnus X-1 we calculated lags in the narrow
energy channels. For our analysis we use publicly available data of Rossi
X-ray Timing Explorer observations P10238 performed between Mar. 26, 1996 and
Mar. 31, 1996 with total exposure time $\sim$70 ksec (we used only
observations with all 5 PCU turned on). We used PCA data in
the ``Generic Binned'' mode, having $\frac{1}{64}$sec$\sim$16 msec time 
resolution in 64 energy channels covering the whole PCA energy band
(B\_16ms\_64M\_0\_249). The lags were calculated for every channel above
3.7 keV, with respect to the count rate in the 2.8-3.7 keV energy
band. The resulting energy dependence of time lag at the frequencies
of the order of 2.5 Hz is shown in Fig.\ref{lag}\footnote{Time
lags at other frequencies in the range from $\sim$0.6 to $\sim$10 Hz have
approximately similar energy dependence (see Fig.\ref{compar})}. For
comparison we plot in the same figure an expected energy dependence of
the lags due to the reflected component, according to equation
(\ref{eq:refl}). Normalization of the expected curve is
arbitrary. Here we used as $D(E)$ the averaged spectrum of the
source during this observation. Reflected component $R(E)$ was calculated
using XSPEC V11 (Arnaud 1996) model {\rm\bf pexrav} of Magdziarz \& Zdziarski
1995 plus a Gaussian at the energy of 6.4 keV. The resulting spectrum has
been convolved with the PCA energy response matrix. Note that expected lags
were calculated using equation (\ref{eq:refl}), i.e. with respect to the 
energy range where contribution of the reflected component was assumed
to be zero. Therefore expected lags do not go to zero at the energy
$<$4 keV. Two curves shown in Fig.\ref{lag} differ in the normalization of
the Gaussian added to the reflected continuum. For upper curve we used
relative normalization of the Gaussian and reflected continuum obtained from
the fit to the total Cygnus X-1 spectrum during observation 10238-01-08-00
(Gilfanov et al., 1999). For the lower curve the normalization of the
Gaussian was chosen such that the equivalent width of the line with respect to
the reflected continuum is $\sim$1 keV. The latter value is characteristic
for an angle averaged reflection from a neutral matter with a normal
abundance of heavy elements. 

One can see from  Fig.\ref{lag} that the lags due to the reflected
component should contain a prominent feature at the energy
$\sim$6.4 keV. We stress again that this feature should be much more
prominent in the lags than in the averaged spectrum, where iron line
provides at most 10\% relative deviation of the observed spectrum from a
power law fit. Clearly no such feature is present in the observed lags. We 
therefore conclude that the observed lags are predominantly not due to
the reflected component. 

One can use the same argument in order to constrain geometry of the
reflecting region. Indeed lack of the iron line feature in the observed
energy dependence of the time lags imply certain constrains on the
combination of the reflected component strength and the duration of the
time delay due to the finite size of the reflector. In the
Fig.\ref{compar} we show the observed energy dependence of the time lags
at three representing frequencies: 0.6, 2.5 and 10 Hz. For comparison
we show in the same plot expected time lags for different geometries
of the reflector and different strength of the reflected
component. 
Four basic models have been considered:
\begin{itemize}
\item {\bf A}: A point isotropic source above the flat disk with the
hole in the middle. The height $h$ of the source above the disk plane is set
to 10$R_g$, the size of the hole $R_{in}$ in the disk is set to
50$R_g$. Here $R_g$ is the gravitational radius for a 10$M_\odot$
black hole. The factor $\Omega/2\pi$, calculated for this geometry,
is $\simeq$ 0.2. 
\item {\bf B}: The same geometry as for {\bf A}, but the strength of
the reflected component was enhanced, so that effective parameter
$\Omega/2\pi$ is 0.54. This is possible if e.g. the source of the
primary emission is anisotropic and more emission is coming towards
the disk than to the observer. Particular value of the normalization of the
reflected component in this model was taken from the fit to the total Cygnus
X-1 spectrum during observation 10238-01-08-00 (Gilfanov et al., 1999).
\item {\bf C}: The same geometry as for {\bf A}, but  for $R_{in}=10
R_g$.The factor $\Omega/2\pi$, calculated for this geometry, 
is $\simeq$ 0.7
\end{itemize}

The transfer functions in the form $\frac{\Omega}{2\pi} T(t)$ for models
{\bf A} and {\bf C} are shown in Fig.\ref{transfer} for two values of the
disk inclination $\theta$ ($0^\circ$ and
$50^\circ$). Model {\bf B} obviously has the same shape of the transfer
function as model {\bf A} and is different only in normalization. For
comparison we also show the transfer function for a concave disk with the
dependence of disk height over radius adopted from Shakura and Sunyaev
(1973). For the frequency range considered here such concave disks do not
differ from the flat disks. 

The lags expected in these models (for inclination $\theta=50^\circ$) are
shown in Fig.\ref{compar}. The model curves were
calculated for the equivalent width of the Fe $K_{\alpha}$ line with respect
to the reflected continuum obtained from the fit of the time averaged source
spectrum. As was mentioned above lack of prominent iron feature in the
energy dependence of the time lags suggests that lags are largely not due
to the reflection. Moreover lags expected in models {\bf B} and {\bf C} at
10 Hz must produce clear feature at 6.4 keV in contradiction to the data. 
Even for model {\bf A} expected lags can be only marginally ``hidden'' in
the observed lags. There are however many uncertainties and simplifications
in the models discussed above. In particular for simplicity we neglected
the dependence of the reflected component spectrum on the incident angle.
Also the surface of the reflector (upper few Thomson optical depths) may not
be smooth and flat, but wrinkled. Furthermore ionization of the reflector
surface will modify the spectrum of the reflected component. Finally all
relativistic effects associated with reflection from the inner regions of
the accretion disk (see e.g. Campana and Stella 1995) have been ignored.
Account for all these effects may modify the normalization and shape of the
energy dependence of the time lags by a factor of few. 

Perhaps the most dubious assumption is the ``linear'' response of the
reflected component to the variations of the illuminating flux. Recent
analysis (e.g. Nayakshin and Kallman 2000) shows that shape of the
reflected component can vary in a complicated fashion in response to
increase of the primary flux. Therefore conservative conclusions of
this section are that i) ``linear'' reflected component,
containing prominent fluorescent line at 6.4 keV, is not responsible 
for the observed lags and ii) some of the source/reflector geometrical
models can be definitely excluded.  

\section{Time lags due to variations of the power law index}
The spectrum of Cygnus X-1 in the 2--20 keV energy band can be
reasonably well approximated by a power law. Deviations, in particular
due to the reflected component, are present, but their amplitude relative
to the power law component is usually less than 10\% in this energy range.
We assume below that at any moment of time the spectrum can be
represented as a power law:
\begin{eqnarray}
S(E,t)=A(t)E^{-\alpha(t)}
\end{eqnarray}
where $A(t)$ is the normalization of the power law as a function of
time and $\alpha(t)$ is the photon index, which is also a function
time. For simplicity we set $\alpha(t)=\alpha_0-\beta(t)$, where
$\alpha_0$ is a photon index of the time averaged spectrum and
$\beta(t)$ is the time variable part of the photon index. Assuming
that $\beta(t)\ll \alpha_0$ the spectrum of the source can be
rewritten as:
\begin{eqnarray}
S(E,t)=A(t)E^{-\alpha_0}(1+\beta(t)lnE)
\end{eqnarray}

\begin{figure}
\plotone{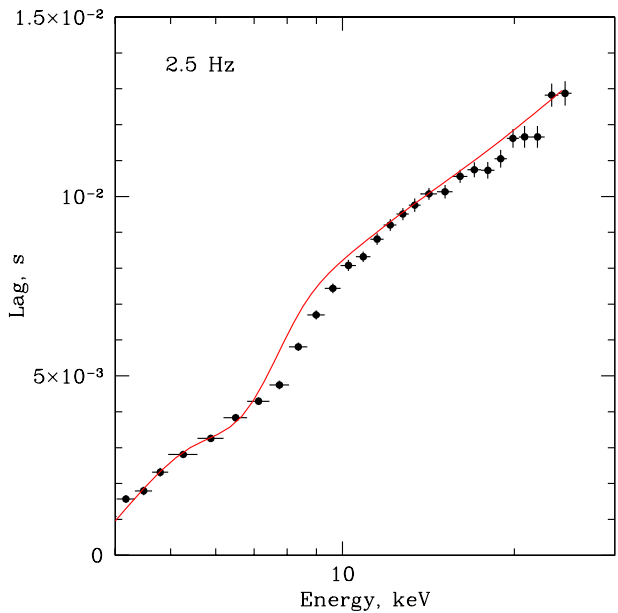}
\caption{Time lag at the frequency of 2.5Hz as a function of
$Log(E)$. See text for the description of the solid line.
\label{f:loge}
}
\end{figure}

The phase lag $\phi$ at a given frequency $f$ can then be written
through the Fourier transforms $\hat{S}^*(E,f)$ of the source light
curves at two different energies $E_1$ and $E_2$ as: 
\begin{eqnarray}
sin\phi=\frac{Im \left [ \hat{S}^*(E_1,f) \hat{S}(E_2,f)\right
] }{|S(E_1,f)||S(E_2,f)|}\approx  \nonumber \\
\frac{Im \left [ (\hat{A}^*+\hat{(\beta A)}^*lnE_1)
(\hat{A}+\hat{(\beta A)}lnE_2)\right ]}{\hat{A}^*\hat{A}} \approx
\nonumber \\
\frac{Im \left [ \hat{A}^*\hat{(\beta A)} \right
]}{\hat{A}^*\hat{A}}\times ln\frac{E_2}{E_1}~~~~~~~~~
\end{eqnarray}
In the above expression we neglected terms of the order of $O(\beta^2)$.
 Thus the assumption that the 
spectrum has a power law shape at any moment of time and the
variations of slope are small automatically implies that phase lags
(and consequently time lags) scales as $ln\frac{E_2}{E_1}$.

\subsection{Universal slope flux correlation?}
It is clear that if variations of slope $\beta(t)$ and normalization
$A(t)$ are random and incoherent then the time lag is zero. If these
two quantities are correlated then nonzero lag may appear. The
simplest assumption would be a linear correlation between slope and
normalization (i.e. between slope and flux):
\begin{eqnarray}
\beta(t)=\gamma\times(A(t)-A_0)
\label{eq:bet}
\end{eqnarray}
where $A_0$ is the normalization of the time averaged spectrum and
$\gamma$ is the coefficient of linear proportionality between
variations of normalization and slope.

The expression for phase lags is then further reduces to:
\begin{eqnarray}
sin\phi=\frac{Im \left [ \hat{A}^*\hat{(A^2)} \right
]}{\hat{A}^* \hat{A}}\times \gamma ln\frac{E_2}{E_1}~~~~~~~~~
\label{eq:gam}
\end{eqnarray}
In this approximation the sign and the amplitude of the lags would
simply reflect properties of the flux variations on different time
scales. Unfortunately the function (\ref{eq:gam}), calculated for the
observed light curves, changes the sign several times within the
frequency range where hard lags are observed. This means that simple
linear correlation between the flux and the slope of the spectrum can be
excluded or that sign of the correlations between flux and slope also
has to change sign in order to keep the sign of (\ref{eq:gam}) the
same at all frequencies. We note here that change of the slope/flux
correlations sign at different time scales was indeed reported for
Cygnus X-1 (Li, Feng \& Chen 1999). Anyway it is clear that simple
linear correlation between the slope and the flux fails to produce
observed hard lags and more complicated models are to be invoked. 

\section{Time lags due to accretion}

In this section we construct simple model which attributes the lags to the
propagation of perturbations  in the accretion flow from the outer to
inner region. As was discussed by Lyubarskii (1997) and Churazov,
Gilfanov and Revnivtsev (2001) strong variability of the X--ray flux
over very broad dynamic range of time scales suggests that different
time scales are introduced to the accretion flow at different
distances from the compact source. One might think e.g. in terms of
the effect on the mass accretion rate of the magneto-hydrodynamic
turbulence, which serves as a source of the viscosity in
the accretion flow through the fluctuating magnetic stresses
(e.g. Balbus and Hawley 1991, Hawley, Gammie and Balbus 1995,
Brandenburg, Nordlund, Stein and Torkelsson 1995). The perturbations
introduced at large distance from the compact object are then
propagated down to the region of main energy release and cause
observed variations of the X--ray flux. If the propagation of the
perturbations in radial direction is due to viscous diffusion then the
very fact that we observe strong variations of the X--ray flux means
that propagation/diffusion time from a given radius $r_0$ is comparable to
or shorter than the time scale of perturbations, introduced to the flow
at this radius $r_0$. Otherwise these perturbations would be completely
washed out before reaching the innermost region.  This in turn means
that lags could naturally appear in such situation if locally emitted
spectrum is a function of radius $r$. Of course energy budget of the
accretion flow far from the innermost region is small, but the
observed lags are also small, of the order of 0.1 radian. We then
tested feasibility of this model using following assumptions: 
\begin{itemize}
\item Statistically independent (incoherent) perturbations are
introduced to the accretion flow at different distances from the
compact source. The characteristic time scale of these perturbations is
a function of radius. 
\item We further assume that the shape of the locally emitted spectra 
depends on radius and it is getting progressively softer with
increase of the distance from the compact object. 
\end{itemize}
The light curve at a given energy E is:
\begin{eqnarray}
L(E,t)=\nonumber \\
\int \int \int  D(r_0,t_0) G(r,r_0,t-t_0) \epsilon (r) S(E,r) dr dr_0 dt_0
\label{eq:lc}
\end{eqnarray}
where $D(r_0,t_0)$ is an initial perturbation introduced to the flow
at radius $r_0$ and time $t_0$, $G(r,r_0,t-t_0)$ is the Green
function, which describes propagation of the perturbations from radius
$r_0$ to $r$, $S(E,r)$ is the shape of the locally emitted spectrum at
the radius $r$ and $\epsilon (r)$ is the total luminosity of the
accretion flow emitted at a given radius. Assumption of statistical
independence of perturbations introduced at different radii implies
that quantities like power density or cross spectra can be calculated
independently for each initial radius and then averaged over range of
initial radii. The Fourier transform $\hat L(E,f)$ of the X--ray light
curve is then:  
\begin{eqnarray}
\hat L(E,f)=\int \hat D(r_0,f) \int  \hat G(r,r_0,f) \epsilon (r) S(E,r)
dr dr_0
\label{eq:flc}
\end{eqnarray}
where $\hat D(r_0,f)$ is the Fourier transform of the perturbations
introduced to the flow at the radius $r_0$, $\hat G(r,r_0,f)$ is the
Fourier transform of the Green function, $f$ is the frequency. From
the above equation it is clear that in this approximation time lags
are solely due to the properties of the Green function and the
dependence of the spectral shape on radius. For our
toy model we choose the simplest possible version with $\epsilon
(r)\propto 1/r^2$ per unit $dr$. In order to calculate phase and time
lags we created two artificial light curves in the ``soft'' and
``hard'' bands assuming simple laws of emissivity in these bands:
\begin{eqnarray}
S(r)=\epsilon (r) S(E_{soft},r)=\frac{1}{r^2}~~~~~~~~~~ \nonumber \\
H(r)=\epsilon (r) S(E_{hard},r)=\frac{1}{r^2}\times h(r)
\label{eq:sh}
\end{eqnarray}
where $h(r)$ is a decreasing function of the radius. 
For simplicity we assume that propagation of the perturbations can be
characterized by the Green function of Lynden-Bell and Pringle (1974)
describing diffusion in the geometrically thin disk
approximation (see also Lyubarskii 1997). I.e.
\begin{eqnarray}
G(r,r_0,t)\propto \frac{r^{-1/4}}{t} exp\left [-\frac{r_0^{1/2l}+r^{1/2l}}{4t} \right ] I_l \left[ \frac{r^{1/4l}r_0^{1/4l}}{2t}\right ]
\label{eq:gf}
\end{eqnarray}
where $I_\nu (x)$ is the Bessel function
of imaginary argument. For the parameter $l$ we use the value of $1/3$
appropriate for a standard $\alpha$ disk with constant ratio
$\frac{H}{r}$, where $H$ is a half thickness of the disk.  In this
equation and below we measure radius $r$ in units of $r_{min}$, which is
supposed to be the characteristic radius of the inner region of the
accretion flow where most of the energy is released. Time is measured
in units of the characteristic diffusion time $k^2\times
r_{min}^{1/2l}$, where $k$ is the effective diffusion coefficient. 

\begin{figure}
\plotone{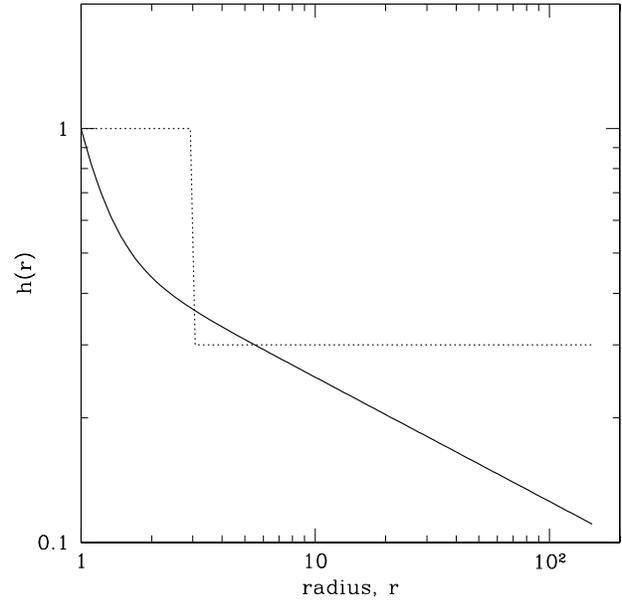}
\caption{Two versions of $h(r)$ function, used in the
simulations. Dotted line shows the $h(r)$ function when the locally
emitted spectrum has the same slope at all distances from the compact
object except for the innermost region, where the spectrum is
harder. Solid line shows the case when spectrum gradually steepens
with distance from the compact object. The amplitude of the $h(r)$
function change affects the absolute value of the time/phase lag.
\label{f:hr}
}
\end{figure}

\begin{figure}
\plotone{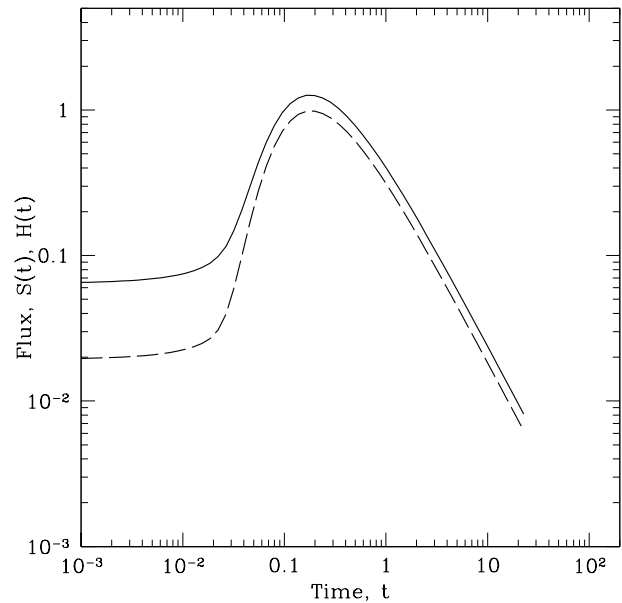}
\caption{Light curves in two energy bands (soft -- solid line, hard --
dashed line) for the $\delta(t)$ perturbation at the initial radius
$r_0=30$. 
\label{f:lcexample}
}
\end{figure}

Two simple versions of the function $h(r)$ (hereafter ``hardness''
function) are shown in
Fig.\ref{f:hr}. The dotted line shows the case when slope of the
locally emitted spectrum is the same at all distances from the compact
object, but it abruptly hardens within innermost region. The solid
line shows the case when locally emitted spectrum gradually hardens as
the distance from the compact object decreases. In Fig.\ref{f:hr}
light curves in the soft and hard energy bands are
shown, assuming $h(r)$ function with the abrupt jump. Here initial
radius $r_0$ was set to 30 and initial perturbation was assumed to be
$\delta$ function in time and radius. Initially locally emitted
spectrum is soft and it hardens when perturbation spreads all the way
down to the minimal radius. After that moment hardness of spectrum
remains practically unchanged. 

\begin{figure}
\plotone{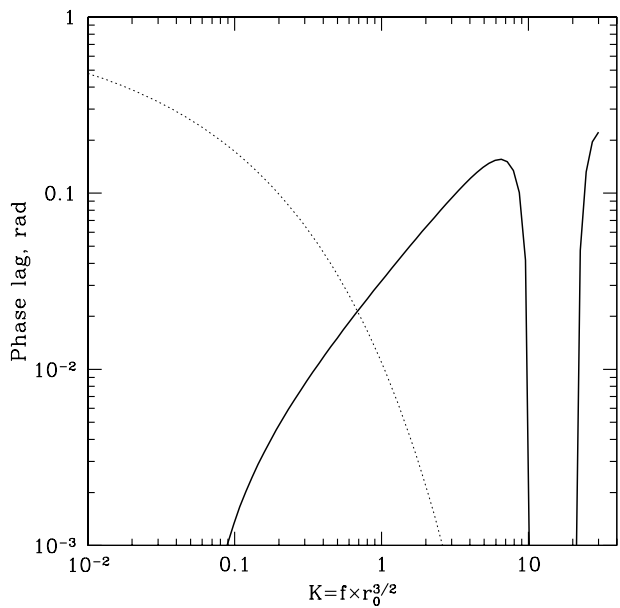}
\caption{Phase lag (solid line) as a function frequency for the perturbations
introduced at fixed initial radius $r_0=30$. Dotted line show the
value of the transfer function power  $| \int  \hat G(r,r_0,f)
\epsilon (r) dr |^2$, which characterizes the suppression of the power
of the initial perturbation at a given frequency, due to diffusive spreading.
\label{f:fixr}
}
\end{figure}

We then calculated phase lags fixing the initial radius $r_0=30$ and
varying the characteristic frequency of the initial perturbation. The
dependence of the phase lag on frequency is shown in the
Fig.\ref{f:fixr}. Solid line in this figure shows the phase lag (in
radians) as a function of frequency expressed in units of inverse
characteristic diffusion time for a given initial radius
$r_{0}^{3/2}$. These lags were calculated for $h(r)$ function 
with a jump (see Fig.\ref{f:hr}), but for the smoother $h(r)$ the
resulting lags are practically the same. Dotted line in the same
figure shows the value of the transfer function power  $| \int  \hat G(r,r_0,f)
\epsilon (r) dr |^2$. It
characterizes to what extent an amplitude of a perturbation at a given
frequency $f$ is 
suppressed in the observed light curve due to diffusive
spreading. Values of the order of $10^{-2}$ imply that ``local''
perturbations at $r_0$ with an amplitude of the order of unity would
result in $\sim$10\% variations in the innermost region where bulk of
energy is released.

\begin{figure}
\plotone{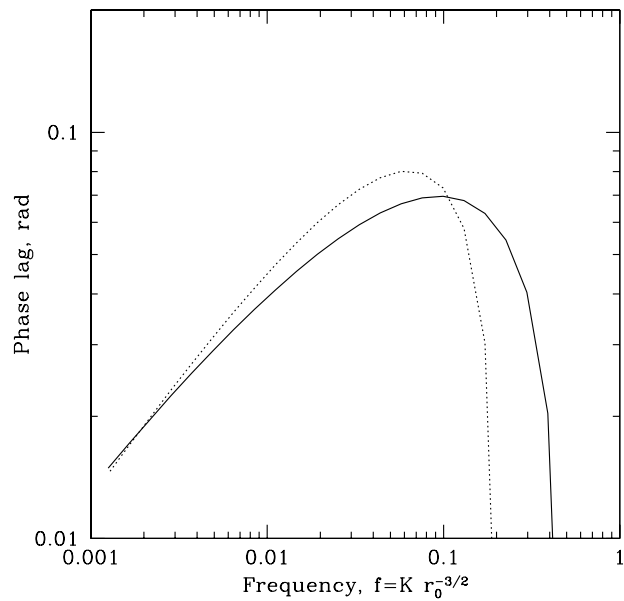}
\caption{Phase lags for two versions of $h(r)$. Frequency, introduced
at radius $r_0$ is assumed to scale as $K\times r_0^{-3/2}$ with
$K=1$.
\label{f:dl}
}
\end{figure}

In order to calculate phase lags associated with the perturbations
generated at different radii we made further simplifying assumption
that at every initial radius $r_0$ instabilities generate single
frequency which scales as a characteristic diffusion time at this
radius\footnote{Note that for the $\alpha$ disk with constant ratio of
the disk height to the radius the characteristic diffusion time and
Keplerian time scales similarly}. I.e. $f(r_0)=K*r_{0}^{-3/2}$, where 
$K$ is a parameter. From Fig.\ref{f:fixr} one would expect the lags to
increase with the increase of $K$ unless $K$ is too large. The resulting
phase lags for two versions of the ``hardness'' function are shown in
the Fig.\ref{f:dl}. The resulting phase lags weakly grows with
frequency and reach $\sim$ 0.1 radian before falling sharply. The
amplitude of the phase lags and the 
dependence of the lags on frequency are thus regulated by two major
parameters: i) the strength of the spectral changes as a function of
radius and ii) the relation between the time scale of  perturbations, locally
generated at radius $r_0$, and diffusion time scales from $r_0$ down to
much smaller radii. We stress that under assumption that at each
radius $r_0$ perturbations with single characteristic frequency are
generated the  phase lags do not depend on the shape of the power
density spectra and are solely determined by the Green function and
the ``hardness'' function. 

\begin{figure}
\plotone{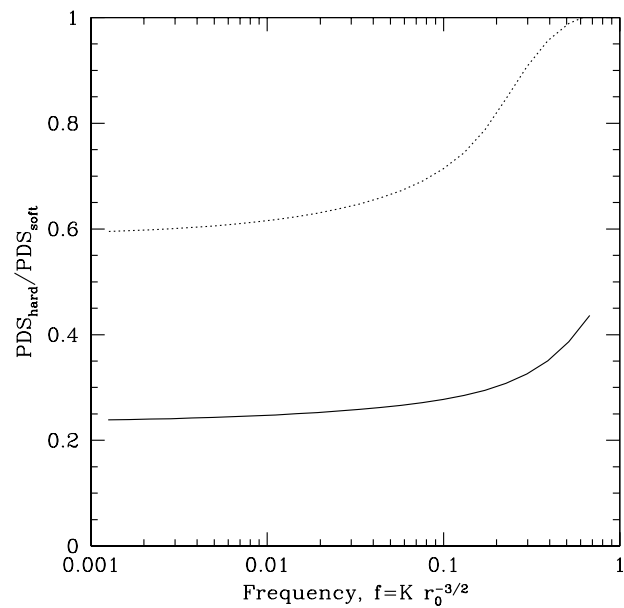}
\caption{Ratio of the power density spectra of the hard and soft light
curves (see eq.\ref{eq:sh}) for two versions of $h(r)$.
\label{f:pdsr}
}
\end{figure}

In the model, considered above, one would expect that power density
spectra in different energy bands should have somewhat different shape
due to the influence of the ``hardness'' function. In the
Fig.\ref{f:pdsr} we show the ratio of the power density spectra in the
``hard'' and ``soft'' bands as a function of frequency. This ratio is
again independent on the shape of the power density spectrum and is
governed by the the Green function and the ``hardness'' function.
One can see that, as expected, the ratio of the power density spectra
increases with frequency. 

\begin{figure}
\plotone{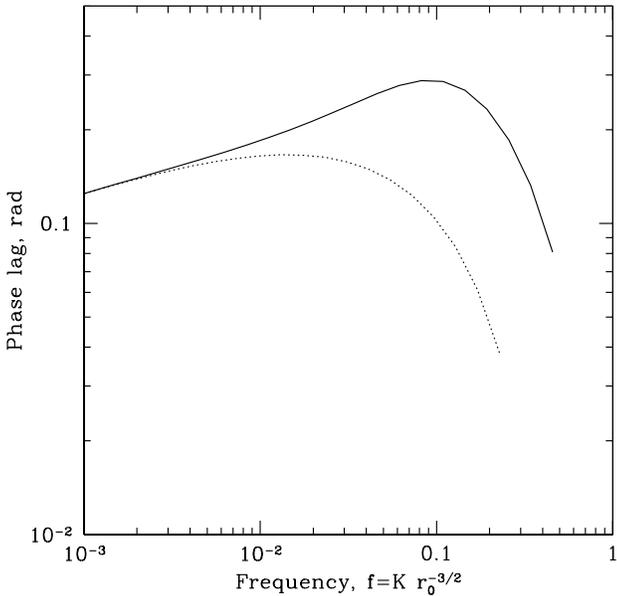}
\caption{Lags for the free fall model for $K=1$ and $K=0.5$.
\label{f:ffl}
}
\end{figure}

For completeness we also calculated time lags using the same choice of
the ``hardness'' function, but instead of diffusion Green function we
substitute Green function appropriate for a free fall from given radius
$r_0$. This is equivalent to the assumption that distinct blobs of
emitting matter are falling towards the compact source with the
velocity $v\propto r^{-1/2}$. The resulting lags are shown in
Fig.\ref{f:ffl}. Since in this case no viscous spreading is present,
the absolute value of the lags is higher for the same choice of the
``hardness'' function and parameter $K$.

Finally we stress that if the locally emitted spectrum at any radii is
a power law then observed lags will show logarithmic dependence on
energy even if the total observed spectrum (integrated over all radii) is not
well represented by a power law as assumed in the previous
section. This follows from our assumption that perturbations generated
at different radii are statistically independent.

\subsection{``Antilags'' due to reflected component}
As we see in Section 2 observed energy dependence of the time lags
contradicts the assumption that hard lags are predominantly produced
by the time delay of the reflected component with respect to the
primary continuum. In particular (see Fig.\ref{f:loge}) instead of a
prominent hump at the energy of the iron fluorescent line the observed
lags seem to be slightly suppressed here. This behavior might hint
that reflected component somehow works in opposite direction and
reduces the lags. We now discuss how such suppression could be
introduced within the frame of the model considered above.

An assumption that locally emitted spectrum
has a power law shape with an index slowly varying with the distance
(so that hardest spectra are near the center) from the compact object
will naturally lead to a hard lags with a logarithmic 
dependence of lags on energy. Let us now assume that locally
emitted spectra also contain a reflected component. Assuming that
overall geometry of the accretion flow resembles a truncated optically
thick disk followed by an optically thin flow one would expect that
the reflection fraction is higher for spectra emitted at larger
distances from the compact object. I.e. softer spectra (emitted at
larger distance from the compact object) have larger reflection
fraction than harder spectra. In our model this means that reflected
component is leading the harder spectrum. In this situation one would
expect slight modification of the logarithmic dependence of lags on
energy due to reflected component. The expected shape of the lags
dependence on energy in this case is shown in Fig.\ref{f:loge} with
the solid line. For simplicity we assumed that lags are due to two
power law spectra with the power law indices of 2.4 and 1.8 and the
reflection fraction of $\sim$1 and $\sim$0.5 respectively. I.e. the
light curve is:
\begin{eqnarray}
L(E,t)=S_1(E)*\delta(t)+S_2(E)*\delta(t-\Delta t)
\label{eq:wlc}
\end{eqnarray}
Reflected components in both spectra were smeared with a $\sim$0.8 keV
Gaussian. In the limit of small frequencies one can easily write an
expression for the time lag as a function of energy for the light
curve described by eq. (\ref{eq:wlc}). Detailed shape of the energy
dependence is affected by strength of the reflection features in both
spectra, amount of smearing applied, relative normalization of both
spectra. One illustrative example, calculated for one particular
choice of these parameters, is shown in Fig.\ref{f:loge} with a solid
curve. This curve is not supposed to closely reproduce the data, but
just to demonstrate that the trend is in right direction. 

\section{Discussion}
\begin{figure}
\plotone{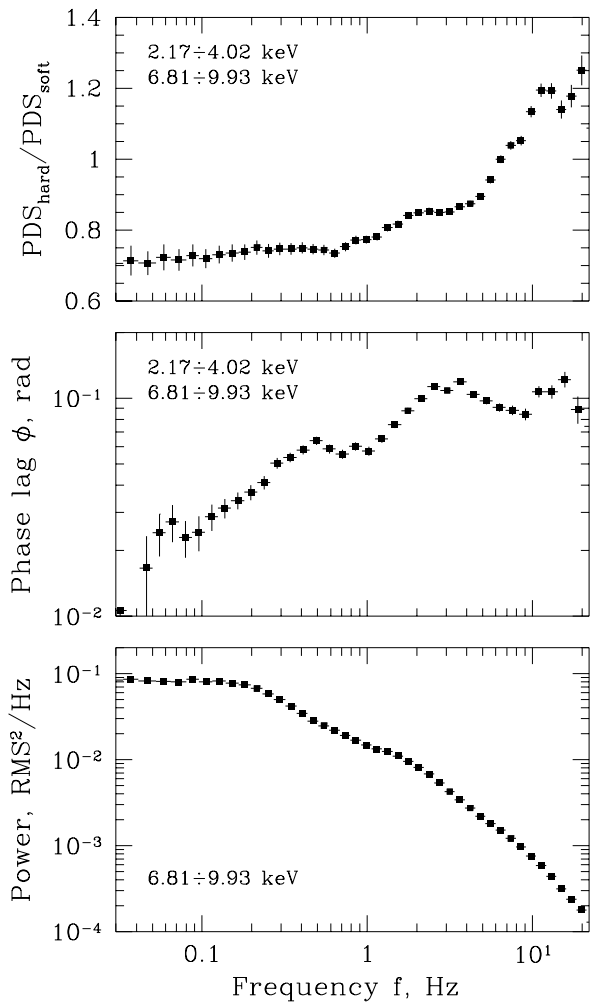}
\caption{Cyg X-1: observed power density spectrum (lower panel), the
phase lags (middle panel)  and ratio of the power density spectra in
the two energy bands (upper panel).
\label{f:cyglags}
}
\end{figure}

The properties of the hard lags, observed in Cyg X-1 can be summaries
as following:
\begin{itemize}
\item In the hard state of Cygnus X-1 lags are observed very
significantly in the frequency range of few $10^{-2}$--few 10
Hz. Phase lag is a slightly growing function of frequency with a maximum
phase lag of $0.1$ rad between 3 and 10 keV (see recent data of Nowak
et al. 1999a, Pottschmidt et al. 2000, also Fig.\ref{f:cyglags}).
\item An energy dependence of the lags is approximately logarithmic
(Miyamoto et al. 1988, Nowak et al. 1999a, see also Fig.\ref{f:loge}) with
the slight, but statistically significant, deviations from this
dependence.
\item The cross correlation function peaks at zero lag (Maccarone et
al. 2000).
\item The coherence function is close to unity (Nowak et al. 1999a), at least
in the middle of the frequency range, where lags are observed with
high significance 
\end{itemize}
Several different branches of models have been suggested as an
explanation for the observed hard lags (see Poutanen 2001a,b for
recent reviews). Our model belongs to the  class of ``propagation''
models as defined by Nowak et al. 1999b. Other models involving
``propagation'' are those of Kato 1989, Nowak et al. 1999b, B\"ottcher
and Liang 1999, Misra 2000, (see also Manmoto et al., 1996). In fact
our version of the model is close 
to the original suggestion of Miyamoto et al. 1988, Miyamoto and
Kitamoto 1989 that {\it 
a clump of matter ... drifts from the outer (soft-X-ray-emitting)
region to the inner (hard-X-ray emitting) region}. 

The model is based on two major assumptions: {\bf a)}
different time scales are introduced to the accretion flow at
different distances from the compact object and propagation time
scales are comparable to the time scales of perturbations and {\bf b)}
locally emitted spectra are power laws with the photon indices smaller in
the innermost region.

The logarithmic energy dependence of the lags follows directly from
the assumption {\bf (b)}. Zero shift of the cross correlation function
and coherence function close to unity naturally appear because most of
the emission is coming from the innermost region and therefore most of
the flux at all energies is released at the same time (see
Fig.\ref{f:lcexample}). The dependence of 
the lags on frequency and maximal values of the lags are tunable
parameters of the model (through the dependence of the photon index on
the radius -- the ``hardness'' function $h(r)$ and through the
relation of the perturbation and propagation time scales). Examples
shown above (see e.g. Fig.\ref{f:dl}) show that with reasonable
assumptions one can approximately reproduce both the dependence of
lags on frequency and the maximal value of the phase lag. 

In addition this model provides natural explanation of the ratio 
of the power density spectra in different energy bands as a function
of frequency (see Fig.\ref{f:pdsr} and \ref{f:cyglags}, upper panel)
and offers a 
qualitative explanation for the deviations of the energy dependence of the lags
from a simple logarithmic law (Fig.\ref{f:loge}). Broadly the model
is consistent with the geometrical model in which the inside truncated
optically thick disk is followed by the optically thin flow. At the
distances larger than the disk truncation radius an optically thin
flow seems to be also present and perhaps has a form of a corona
sandwiching the disk (see arguments in Churazov et al. 2001). The
latter assumption allows one to extend this model to the soft state of
Cyg X-1 without any modifications.

Our assumption that each frequency in the observed flux variability is
associated with some particular radius is of course a gross
oversimplification. However due to suppression of the high frequency
variations in the course of diffusion such situation may in practice
materialize in the accretion flow if the times scales of locally
induced perturbations at given radius are high, e.g. higher than the
diffusion frequency (Churazov et al. 2001). From Fig.\ref{f:cyglags} it is
clear that there are several ``humps'' in the power density spectrum
of Cyg X-1. Assuming that this is an indication that perturbations
arising at some particular radius  of the accretion flow are dominating at
a range of frequencies one would then expect the phase lags to rise
faster with frequency (compare Fig.\ref{f:dl} and \ref{f:fixr}). It seems that
such behavior is indeed observed as a distinct regions of fast phase
lag rise (or time lag shoulders -- see Nowak et al. 1999a,b) in the
Cyg X-1 data (Fig.\ref{f:cyglags}).

Particular emission mechanism is not very important as long as only
the origin of lags is considered. E.g. emission could be due to
magnetic flares as in the picture of Galeev, Rosner, Viana 1979 or 
Poutanen and Fabian 1999. In our model flares spectra need not to
evolve (from soft to hard) themselves. They could be short and could occur
at different distances from the compact object. The observed
variability could then be not due to individual flares but rather due
to change of the number of active flares at a given moment of time. 
The model requires however that the spectra of a more distant flares
to be softer than those of the flares in the innermost region of the
accretion flow.

\section{Conclusions}

We show that some properties of the black hole candidates spectral
variability in the X--ray band (in particular time lags) can be explained by
a simple phenomenological model. This model assumes that {\bf a)}
different time scales are introduced to the accretion flow at
different distances from the compact object and propagation time
scales are comparable to the time scales of perturbations and {\bf b)}
locally emitted spectra are power laws with the photon indices smaller
in the innermost region. 

We also show that energy dependence of the time lags excludes the
possibility that lags (in the 0.1 -- 10 Hz range) are predominantly
due to the delay of the reflected component with respect to the
illuminating continuum. This conclusion is derived under assumption of
a ``standard'' neutral reflected component which linearly responds to
the variations of the illuminating flux.

\section{Acknowledgements} 
This research has made use of data obtained 
through the High Energy Astrophysics Science Archive Research Center
Online Service, provided by the NASA/Goddard Space Flight Center.
O.Kotov acknowedges partial support by RFBR grant 00-15-96649.

\label{lastpage}


\begin{thebibliography}{}
\bibitem[Arnaud 1996]{ar96} Arnaud K.A. 1996,
         Astronomical Data Analysis Software and Systems V,
         eds. Jacoby G. and Barnes J., p17, ASP Conf. Series volume
         101. 
\bibitem[Balbus \& Hawley(1991)]{1991ApJ...376..214B} Balbus, S.\ A.\ \& 
Hawley, J.\ F.\ 1991, \apj, 376, 214 
\bibitem[Basko, Sunyaev, \& Titarchuk(1974)]{1974A&A....31..249B} Basko, 
M.\ M., Sunyaev, R.\ A., \& Titarchuk, L.\ G.\ 1974, \aap, 31, 249 
\bibitem[B{\"o}ttcher \& Liang(1999)]{1999ApJ...511L..37B} B{\"o}ttcher, 
M.\ \& Liang, E.\ P.\ 1999, \apjl, 511, L37 
\bibitem[Brandenburg, Nordlund, Stein and Torkelsson 
(1995)]{1995ApJ...446..741B} Brandenburg, A., Nordlund, A., Stein, R.\ F.\ 
and Torkelsson, U.\ 1995, \apj, 446, 741
\bibitem[Campana \& Stella(1995)]{1995MNRAS.272..585C} Campana, S.\ \& 
Stella, L.\ 1995, \mnras, 272, 585 
\bibitem[Churazov, Gilfanov, \& Revnivtsev(2001)]{2001MNRAS.321..759C} 
Churazov, E., Gilfanov, M., \& Revnivtsev, M.\ 2001, \mnras, 321, 759 
\bibitem[Esin, McClintock, \& Narayan(1997)]{1997ApJ...489..865E} Esin, A.\ 
A., McClintock, J.\ E., \& Narayan, R.\ 1997, \apj, 489, 865 
\bibitem[Galeev, Rosner and Vaiana (1979)]{1979ApJ...229..318G} Galeev, A. 
A., Rosner, R. and Vaiana, G. S. 1979, \apj, 229, 318 
\bibitem[George \& Fabian(1991)]{1991MNRAS.249..352G} George, I.\ M.\ \& 
Fabian, A.\ C.\ 1991, \mnras, 249, 352 
\bibitem[Gilfanov, Churazov, \& Revnivtsev(2000)]{2000MNRAS.316..923G} 
Gilfanov, M., Churazov, E., \& Revnivtsev, M.\ 2000, \mnras, 316, 923 
\bibitem[Gilfanov, Churazov, \& Revnivtsev(1999)]{1999A&A...352..182G} 
Gilfanov, M., Churazov, E., \& Revnivtsev, M.\ 1999, \aap, 352, 182 
\bibitem[Hawley, Gammie and Balbus (1995)]{1995ApJ...440..742H} Hawley, J.\ 
F., Gammie, C.\ F.\ and Balbus, S.\ A.\ 1995, \apj, 440, 742 
\bibitem[Kato(1989)]{1989PASJ...41..745K} Kato, S.\ 1989, \pasj, 41, 745 
\bibitem[Li, Feng, \& Chen(1999)]{1999ApJ...521..789L} Li, T.\ P., Feng, 
Y.\ X., \& Chen, L.\ 1999, \apj, 521, 789 
\bibitem[Lynden-Bell and Pringle (1974)]{1974MNRAS.168..603L} Lynden-Bell 
D. and Pringle J. E., 1974, \mnras, 168, 603 
\bibitem[Lyubarskii 1997]{lyu97} Lyubarskii, Y. E., 1997, \mnras, 292, 679 
\bibitem[Maccarone, Coppi, \& Poutanen(2000)]{2000ApJ...537L.107M} 
Maccarone, T.\ J., Coppi, P.\ S., \& Poutanen, J.\ 2000, \apjl, 537, L107 
\bibitem[Magdziarz \& Zdziarski(1995)]{1995MNRAS.273..837M} Magdziarz, P.\ 
\& Zdziarski, A.\ A.\ 1995, \mnras, 273, 837 
\bibitem[Manmoto et al.(1996)]{1996ApJ...464L.135M} Manmoto, T., Takeuchi, 
M., Mineshige, S., Matsumoto, R., \& Negoro, H.\ 1996, \apjl, 464,
L135 
\bibitem[Misra(2000)]{2000ApJ...529L..95M} Misra, R.\ 2000, \apjl, 529, L95 
\bibitem[Miyamoto, Kitamoto, Mitsuda, \& Dotani(1988)]{1988Natur.336..450M} 
Miyamoto, S., Kitamoto, S., Mitsuda, K., \& Dotani, T.\ 1988, \nat,
336, 450 
\bibitem[Miyamoto \& Kitamoto(1989)]{1989Natur.342..773M} Miyamoto, S.\ \& 
Kitamoto, S.\ 1989, \nat, 342, 773 
\bibitem[Nayakshin \& Kallman(2001)]{2001ApJ...546..406N} Nayakshin, S.\ \& 
Kallman, T.\ R.\ 2001, \apj, 546, 406 
\bibitem[Nolan et al.(1981)]{1981ApJ...246..494N} Nolan, P.\ L.\ et al.\ 
1981, \apj, 246, 494 
\bibitem[Nowak et al.(1999)]{1999ApJ...515..726N} Nowak, M.\ A., Wilms, J.\ 
;., Vaughan, B.\ A., Dove, J.\ B., \& Begelman, M.\ C.\ 1999b, \apj, 515, 726 
\bibitem[Nowak et al.(1999)]{1999ApJ...510..874N} Nowak, M.\ A., Vaughan, 
B.\ A., Wilms, J.\ ;., Dove, J.\ B., \& Begelman, M.\ C.\ 1999a, \apj, 510, 
874 
\bibitem[Pottschmidt et al.(2000)]{2000A&A...357L..17P} Pottschmidt, K., 
Wilms, J., Nowak, M.\ A., Heindl, W.\ A., Smith, D.\ M., \& Staubert, R.\ 
2000, \aap, 357, L17 
\bibitem[Poutanen \& Fabian(1999)]{1999MNRAS.306L..31P} Poutanen, J.\ \& 
Fabian, A.\ C.\ 1999, \mnras, 306, L31 
\bibitem[Poutanen(2001a)]{po2001a} Poutanen, J.\ 2001, 
X-ray Astronomy 1999, Stellar Endpoints, AGN and the Diffuse Background,
eds. G. Malaguti,  G. Palumbo,  and N. White, Gordon \& Breach, Singapore,  in press
\bibitem[Poutanen(2001b)]{2001arxt.confE..25P} Poutanen, J.\ 2001, 
Advances in Space Research, accepted
\bibitem[Priedhorsky et al.(1979)]{1979ApJ...233..350P} Priedhorsky, W., 
Garmire, G.\ P., Rothschild, R., Boldt, E., Serlemitsos, P., \& Holt, S.\ 
1979, \apj, 233, 350 
\bibitem[Revnivtsev, Gilfanov, \& Churazov(1999)]{1999A&A...347L..23R} 
Revnivtsev, M., Gilfanov, M., \& Churazov, E.\ 1999, \aap, 347, L23 
\bibitem[Shakura and Sunyaev 1973]{ss73} Shakura N.,
Sunyaev R., 1973, \aap, 24, 337
\bibitem[Sunyaev and Tr\"{u}mper 1979]{st79} Sunyaev R., 
Truemper J., 1979, \nat, 279, 506
\bibitem[Klis 1994]{kl94} van der Klis M., 1994, \apjs, 92, 511
\bibitem[Zdziarski, Lubinski, \& Smith(1999)]{1999MNRAS.303L..11Z} 
Zdziarski, A.\ A., Lubinski, P., \& Smith, D.\ A.\ 1999, \mnras, 303,
L11 
\end{thebibliography}
\end{document}